\documentclass[aps,prl,twocolumn,10pt,showpacs,preprintnumbers,superscriptaddress,amsmath,amssymb]{revtex4-1} 
\usepackage{graphicx}% Include figure files
\usepackage{dcolumn}% Align table columns on decimal point
\usepackage{bm, times, amsmath, amssymb,color}
\begin{document}

\title{Active Brownian Motion in Threshold Distribution of a Coulomb Blockade Model}

\author{Takayuki Narumi}
\email{narumi@athena.ap.kyushu-u.ac.jp}
\affiliation{Department of Applied Quantum Physics and Nuclear Engineering, Kyushu University, Fukuoka 819-0395, Japan}
\author{Masaru Suzuki}
\affiliation{Department of Applied Quantum Physics and Nuclear Engineering, Kyushu University, Fukuoka 819-0395, Japan}
\author{Yoshiki Hidaka}
\affiliation{Department of Applied Quantum Physics and Nuclear Engineering, Kyushu University, Fukuoka 819-0395, Japan}
\author{Tetsuya Asai}
\affiliation{Graduate School of Information Science and Technology, Hokkaido University, Sapporo 060-0814, Japan}
\author{Shoichi Kai}
\affiliation{Department of Applied Quantum Physics and Nuclear Engineering, Kyushu University, Fukuoka 819-0395, Japan}

%\date{}% It is always \today, today,

\begin{abstract}
Randomly-distributed offset charges affect the nonlinear current--voltage property 
via the fluctuation of the threshold voltage of Coulomb blockade arrays.
We analytically derive the distribution of the threshold voltage
for a model of one-dimensional locally-coupled Coulomb blockade arrays,
and propose a general relationship between conductance and the distribution.
In addition, we show the distribution for a long array is equivalent to
the distribution of the number of upward steps for aligned objects of different height.
The distribution satisfies a novel Fokker--Planck equation corresponding to active Brownian motion.
The feature of the distribution is clarified by comparing it with the Wigner and Ornstein-Uhlenbeck processes.
It is not restricted to the Coulomb blockade model, but instructive in statistical physics generally.
\end{abstract}

%73.23.Hk	Coulomb blockade; single-electron tunneling
%05.10.Gg	Stochastic analysis methods (Fokker-Planck, Langevin, etc.)
%02.50.Ng	Distribution theory and Monte Carlo studies
%71.23.An	Electronic structure of disordered solids: Theories and models; localized states

%\keywords{Coulomb blockade, threshold, stochastic process, Fokker-Planck equation, upward-step distribution, surface}
\maketitle

%%%%%%%%%%%%%%%%%%%%%%
{\it Introduction}.---%
Nonlinear phenomena and threshold behaviors are observed in many disordered systems \cite{refset_disorder}. 
A Coulomb blockade (CB) \cite{refset_CB} is one such example
for which characteristically nonlinear current--voltage ($I$--$V$) behavior occurs 
above a threshold voltage $V_{\text{th}}$.
Specifically, CB is the increased resistance 
at low bias voltage of an electronic device having a low-capacitance tunnel junction, 
the thin insulating barrier that lies between two electrodes 
across which electrons tunnel quantum mechanically. 
Owing to CB, the conductance of the device is not constant at low voltage, 
and no current flows below $V_{\text{th}}$.

Studies have explicitly considered types of disorder
and clarified that disorder affects transport phenomena \cite{Middleton1993, Parthasarathy2001, Reichhardt2003, Bascones2008, Suvakov2010}.
Middleton and Wingreen (MW) considered the charge disorder 
that originates from impurities of a substrate \cite{Middleton1993}.
The threshold voltage is sensitive to this charge disorder.
The distribution of $V_{\text{th}}$ has never been derived,
although MW have discussed the mean value and variance \cite{Middleton1993, Bascones2008}.

In this Letter, we focus on the threshold distribution (TD) 
as it leads to understanding the nonlinearity in $I$--$V$ response;
we show that the conductance is represented by the cumulative distribution of $V_{\text{th}}$.
We find an analytic expression for the TD
for a one-dimensional (1D) locally coupled CB array.
In addition, we reveal that the TD in the long-array limit is equivalent to
the distribution for the number of upward steps for aligned objects of different height.
The distribution satisfies a novel Fokker--Planck equation corresponding to active Brownian motion \cite{Schweitzer1997};
i.e.,~overdamped motion of a Brownian particle in a harmonic potential that spreads with time.
This characteristic of the distribution is quite instructive in the field of statistical physics.

%%%%%%
\begin{figure}[!t]
\begin{center}
\includegraphics[width=70mm]{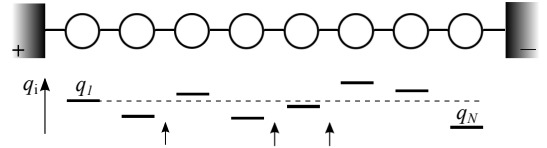}
\end{center}
\caption{
A configuration of a 1D array with $N=8$ (upper) 
and a distribution of the offset charge $q_{\text{i}}$ (bottom).
In the upper figure, a circle represents a Coulomb island
and the connecting line a tunneling junction with capacitance $C$.
The array is sandwiched between positive ($+$) and negative ($-$) electrodes.
Each island connects to a gate electrode, omitted in the figure,
with capacitance $C_{g}$.
In the bottom figure, an arrow indicates an upward step,
which means $q_{\text{l}}<q_{\text{l}+1}$ ($\text{l}\ \in\ \{1,\dots,N-1\}$).
In this distribution, there are three upward steps and four offset charges less than $q_{1}$.
}
\label{fig:1Darray}
\end{figure}

%%%%%%%%%%%%%%%%%%%%%%
{\it Model}.---%
We employ the model proposed by MW \cite{Middleton1993},
in which there are $N$ aligned Coulomb islands, 
constituting the minimum units of charge storage (Fig.~\ref{fig:1Darray}).
We consider that the gate capacitance $C_{g}$ is much greater 
than the island--island and island--electrode capacitances $C$.
In general, interactions such as electron--electron and spin--coupling 
play an important role in evolving the nonlinear $I$--$V$ behavior \cite{Reichhardt2003, Jiang2004}.
However, such interactions are not dominant if $C/C_{g}\ll 1$ corresponding to the so-called locally-coupled CB.
Compared with several theoretical approaches 
such as the density functional theory \cite{Jiang2004, refset_DFT}
and the random matrix theory \cite{refset_RMT} ,
the model we employ is classical and the simplest demonstrating CB.
Many experimental features though can be explained by this model \cite{Kurdak1998, Parthasarathy2001, Son2010, Noda2010, Joung2011},
and theoretical work is still continuing even now \cite{Bascones2008, Suvakov2010, Narumi2011}.
More significantly, some results obtained in this Letter are not solely restricted to this model.

The voltages of the negative and gate electrodes are set to zero, 
and the bias voltage is thus equivalent to the voltage $\Phi_{+}$ of the positive electrode.
Let $Q_{\text{i}}$ denote the charge of the i-th island; $\text{i} \in \{1,\dots, N\}$.
The charge is represented as $Q_{\text{i}}=n_{e}e+q_{\text{i}}$,
where $n_{e}$ denotes an integer, $e$ the elementary charge, 
and $q_{\text{i}}$ the offset charge arise from an impurity.
The offset charges are given by uniform random numbers in [$-e/2, e/2$],
and remain constant over time.
The offset charges just indicate the non-integral part of each charge,
i.e., the uniform distribution for $q_{\text{i}}$ is equivalent to arbitrary distributions of offset charges.

The total energy $E$ of the system is written as \cite{Geigenmuller1989}
\begin{equation} E=\frac{1}{2}\sum_{\text{i, j}}Q_{\text{i}}M^{-1}_{\text{ij}}Q_{\text{j}}+C\Phi_{+}\sum_{\text{i, j}}Q_{\text{i}}M^{-1}_{\text{ij}}+Q_{+}\Phi_{+}, \label{total_energy} \end{equation}
where $Q_{+}$ denotes the charge of the positive electrode.
$M_{\text{ij}}$ denotes the capacitance matrix;
for 1D simple arrays, 
$M_{\text{ij}}=C_{g}+2C$ for $\text{i}=\text{j}$, 
$M_{\text{ij}}=-C$ for $|\text{i} - \text{j}|=1$, 
and $M_{\text{ij}}=0$ otherwise.
The system evolves such that $E$ decreases.
To take the most probable path of evolution,
we transfer an electron to another island 
and calculate the energy change $\Delta E_{\text{i}^{\prime}\to \text{j}^{\prime}}$ for all possible tunneling paths,
where $ \{ \text{i}^{\prime}, \text{j}^{\prime} \} \in \{ 1,\dots, N,+,- \}$. 
In simulations (e.g., \cite{Middleton1993, Suvakov2010, Narumi2011}), 
each tunneling time, which is proportional to the change in energy for $T=0$ \cite{Likharev1986}, is calculated,
and the shortest tunneling time is thus employed for the time evolution increments.
In the rest of the paper, we work in dimensionless units 
whereby the charge is scaled by $e$, the voltage by $e/C_{g}$, and the energy by $e^{2}/C_{g}$.

%%%%%%%%%%%%%%%%%%%%%%
$V_{\text{th}}$ {\it as a function of }$q_{1}$.---%
As a simple example, 
let us consider an array with $N=2$ 
and describe $V_{\text{th}}$ as a function of offset charges.
There are six possible paths; 
however, it is sufficient to consider $\Delta E_{1\to +}$, $\Delta E_{2\to 1}$, and $\Delta E_{-\to 2}$ for $\Phi_{+}>0$.
Note that the paths in the reverse direction should be considered when $\Phi_{+}<0$.
In the limit $C/C_{g} \to 0$,
\begin{subequations}
\label{inequility}
\begin{eqnarray} 
\Delta E_{1\to +} < 0 & ~~\Leftrightarrow~~ & \Phi_{+} > Q_{1}+1/2, \label{1_to_p} \\
\Delta E_{2\to 1} < 0 &~~\Leftrightarrow~~ & Q_{1}-Q_{2} > 1, \label{2_to_1} \\
\Delta E_{-\to 2} < 0 & ~~\Leftrightarrow~~ & Q_{2} > 1/2. \label{n_to_2} \end{eqnarray}
\end{subequations}
If all energy changes are greater than zero, 
no electrons get transferred; i.e., blockading occurs.
Equation (\ref{2_to_1}) suggests that it is effective to separately consider 
charge-offset conditions $q_{1} > q_{2}$ (no upward steps) and $q_{1} < q_{2}$ (an upward step).
As $\Phi_{+}$ increases quasi-statically, under the former condition, 
Eq.~(\ref{1_to_p}) is satisfied above $\Phi_{+}=q_{1}+1/2$,
and an electron then is transferred from island 1 to the positive electrode.
Thus, Eq.~(\ref{2_to_1}) and subsequently Eq.~(\ref{n_to_2}) are satisfied.
Afterward, Eq.~(\ref{1_to_p}) is again satisfied.
This cycle consequently gets repeated; 
i.e., the current flows between the positive and negative electrodes in a steady state above $\Phi_{+}>V_{\text{th}}=q_{1}+1/2$.
In contrast, in the latter case, even if Eq.~(\ref{1_to_p}) is satisfied and an electron moves from island 1 to the positive electrode,
$\Delta E_{2\to 1}$ remains greater than zero because $q_{1}<q_{2}$.
For $\Delta E_{1\to +}$ and $\Delta E_{2\to 1}$ to be less than zero, 
$\Phi_{+}$ has to be increased to $q_{1}+3/2$, 
and a steady-state current then flows; 
i.e., the voltage threshold is $V_{\text{th}}=q_{1}+3/2$.
The above argument holds, without loss of generality, to arbitrary $N$; i.e.,
\begin{subequations}
\label{eq_segment}
\begin{eqnarray} V_{\text{th}}(q_{1}, n) & =  & q_{1}+n-1/2 ~~~ (-1/2\le q_{1} \le 1/2)~~~ \label{Vth_segment} \\ 
\Leftrightarrow~~q_{1}(V_{\text{th}},n) & = & V_{\text{th}}-n+1/2 ~~~ (n-1 \le V_{\text{th}}\le n)~~~ \label{q1_segment} \end{eqnarray}
\end{subequations}
where $n-1$ indicates the number of upward steps; $1\le n \le N$.
The threshold depends only on $q_{1}$ and $n$; 
i.e., the magnitudes of the offset charges between neighboring islands is renormalized to $n$.

%%%%%%%
\begin{figure*}[htdp]
\begin{center}
\includegraphics[width=160mm]{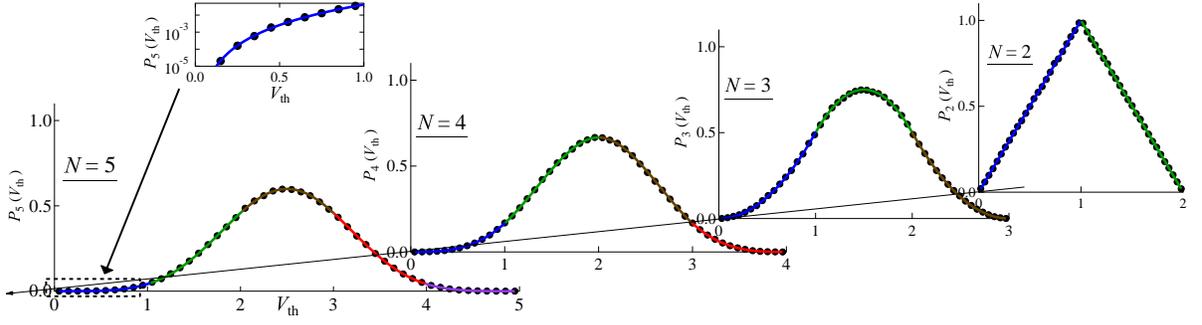}
\end{center}
\caption{(Color online) 
Plot of TDs of $V_{\text{th}}$ for $N=2, 3, 4$, and $5$ (from right to left).
A filled circle represents a simulation result
and a colored line the segmented TD $P^{(n)}_{N}(V_{\text{th}})$ obtained analytically 
for $n=1$ (blue), $n=2$ (green), $n=3$ (brown), $n=4$ (red), and $n=5$ (purple).
$P^{(n)}_{N}(V_{\text{th}})$ for $N=2$ is expressed by Eq.~(\ref{TD_N002}) and expressions for $N=3,4,$ and $5$ are given in the supplement \cite{suppl_01}.
The inset is a close-up (semi-log plot) of the first segment for $N=5$.
The simulation used $10^{6}$ different initial distributions of the offset charges.
}
\label{fig:TD_layout}
\end{figure*}

%%%%%%%%%%%%%%%%%%%%%%
{\it Threshold distribution}.---%
Equation (\ref{Vth_segment}) suggests that the charge-offset analysis based on $q_{1}$ is appropriate.
In addition, Eq.~(\ref{q1_segment}) suggests that the region $0\le V_{\text{th}} \le N$ should be divided into $N$ equally-spaced segments.
Thus, the $n$-th segmented TD for the $N$-island array is expressed as
\begin{equation} P^{(n)}_{N}(V_{\text{th}}) = \sum_{k=0}^{N-1} U_{N}(n|k)\Pi_{N}(k)~~~(n-1 \le V_{\text{th}}\le n), \label{Prob_P} \end{equation}
where $U_{N}(n|k)$ denotes the conditional probability that there are $n-1$ upward steps 
if there are $k$ offset charges less than $q_{\text{h}^{\prime}}$.
Note that $U_{N}(n|k)$ does not depend on $V_{\text{th}}$.
Here, since $q_{1}$ is the basis for analyzing the offset charges, 
we should select $\text{h}^{\prime}=1$.
$\Pi_{N}(k)$ denotes the probability that there are $k$ offset charges less than $q_{1}$, 
and is expressed as 
\begin{equation} \Pi_{N}(k) = \left(\begin{array}{c} N-1 \\ k \end{array}\right) p_{\text{L}}^{k} \  p_{\text{G}}^{N-1-k}, \label{Prob_Pi}\end{equation}
where $p_{\text{G}}$ and $p_{\text{L}}$ are the probabilities of $q_{\text{h}} > q_{1}$ and $q_{\text{h}} < q_{1}$, respectively, 
and $\text{h} \in \{2,\dots,N\}$.
Note that $p_{\text{G}} = 1/2-q_{1}$ and $p_{\text{L}} = 1/2+q_{1}$.

One can obtain $U_{2}(1|0)=U_{2}(2|1)=0$ and $U_{2}(1|1)=U_{2}(2|0)=1$, and then,
\begin{equation} P^{(1)}_{2}(V_{\text{th}}) = V_{\text{th}}~,~~P^{(2)}_{2}(V_{\text{th}})=2-V_{\text{th}}. \label{TD_N002} \end{equation}
Using the same procedure, 
we obtain the entire TD $P_{N}(V_{\text{th}})$ for arbitrary $N$
as the joining of the segmented TDs $P^{(n)}_{N}(V_{\text{th}})$ \cite{suppl_01}.
As shown in Fig.~\ref{fig:TD_layout}, simulation results are correctly described without fitting parameters.
It is clear that, for arbitrary $N$, each segmented TD is represented as an ($N-1$)-degree polynomial of $V_{\text{th}}$ 
because of the term ${p_{L}}^{k}  {p_{G}}^{N-1-k}$.
For small $N$ (in particular, $N=2$ in Fig.~\ref{fig:TD_layout}), the distributions have strange shape
which might be a consequence of model-dependent behavior.
In more realistic cases, other physical effects such as electrode shape should be taken into account.

%%%%%%%%%%%%%%%%%%%%%%
{\it Distribution of upward steps}.---%
The conditional probability $U_{N}(n|k)$ determines the TD for arbitrary $N$.
However, in practice, it is difficult to obtain $U_{N}(n|k)$ for large $N$.
To investigate the TD for large $N$, we focus on the intersections of the segmented TDs.
In particular, we focus on the right edge of each segment; i.e., $V_{\text{th}}=n$.
Since $(p_{L}, p_{G})=(1,0)$ at the right edge,
Eq.~(\ref{Prob_P}) reduces to
\begin{equation} P^{(n)}_{N}(V_{\text{th}}) = U_{N}(n|N-1)=:Y(n, N) ~~~(\text{at}~V_{\text{th}}=n). \label{def_Y} \end{equation}
Therefore, our problem results in obtaining $Y(n, N)$ that indicates the probability 
in the case of $n-1$ upward steps for $N-1$ aligned objects (i.e.,~$q_{2},\dots, q_{N}$) of different height.
Since none of the specific features of the model are used,
the discussion in the rest of this section is not limited to CB 
but has applicability to statistical physics generally.

\begin{figure}[!t]
\begin{center}
\includegraphics[width=70mm]{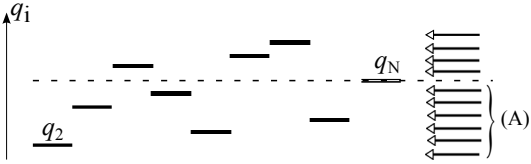}
\end{center}
\caption{
An example of the distribution, where there are five upward steps ($n=6$) and five offset charges less than $q_{N}$ ($k=5$).
The arrows at right indicate the possible location of $q_{N+1}$.
There are $N$ arrows in total, with $k+1$ arrows belonging to (A).
If an arrow is chosen from (A), no increase in upward steps occurs.
}
\label{fig:explain}
\end{figure}

We consider the probability that the number of upward steps for $N+1$ different heights is the same as that for $N$ different heights.
According to Fig.~\ref{fig:explain}, the probability is expressed by $\left<k+1\right>/N$,
where the brackets $\left<\cdot\right>$ indicate the average for 
\begin{equation} D^{(n)}_{N-1}(k):=U_{N-1}(n|k)\left/\sum_{k=0}^{N-2}U_{N-1}(n|k)\right. .\end{equation}
$D^{(n)}_{N-1}(k)$ denotes the probability that there are $k$ offset charges less than $q_{\text{h}^{\prime}}$
if there are $n-1$ upward steps in $N-1$ offset charges,
where the basis for analyzing offset charges is $q_{N}$, i.e., $\text{h}^{\prime}=N$.
Although a mathematical proof has yet to be given,
the probability $\left<k+1\right>/N$ is expected to be $n/N$ \cite{footnote:confirmation}.
This expectation is understandable qualitatively as follows.
If there are already many upward steps (i.e., large $n$), 
then $q_{N}$ tends to be greater than other offset charges ($q_{2},\dots,q_{N-1}$).
Thus, the probability tends to increase with increasing $n$.
With this expectation, the recurrence formula for $Y(n,N)$ is obtained as
\begin{equation} Y(n,N+1)=\frac{n}{N}Y(n, N)+\frac{N-(n-1)}{N}Y(n-1,N). \label{recurrence} \end{equation}

For later discussion, we introduce both a fictive field $x=n-N/2$ and time $t=N$.
Note that $x$ and $t$ do not indicate electron motions, 
but are just changes in variables \cite{suppl_02}.
By defining $Z(x,t):=Y(x+N/2,N)$, Eq.~(\ref{recurrence}) reduces to
\begin{equation} Z(x,t+1)=\left(\frac{1}{2}+\frac{x_{+}}{t}\right)Z\left(x_{+},t\right)+\left(\frac{1}{2}-\frac{x_{-}}{t}\right)Z\left(x_{-},t \right), \end{equation}
where $x_{\pm}=x\pm 1/2$.
In the continuous limit, a partial differential equation is obtained
\begin{equation} \frac{\partial Z(x,t)}{\partial t} = \frac{\partial}{\partial x}\left[\frac{x}{t}Z(x,t) \right]+D\frac{\partial^{2} Z(x,t)}{\partial x^{2}}, \label{PDE_Z} \end{equation}
for which $D=1/8$ describes locally-coupled CB.
The first term of r.h.s.~depends explicitly on time,
so that the equation is classified as related to a time-dependent Ornstein--Uhlenbeck (OU) process \cite{Gardiner2004}.
The differential equation is equivalent to the Fokker--Planck equation corresponding to active Brownian motion \cite{Schweitzer1997};
i.e.,~overdamped motion of a Brownian particle in a harmonic potential $\phi(X,t) = X^{2}/2t$, represented as
\begin{equation}  \frac{{\rm d}X}{{\rm d}t} = - \frac{\partial}{\partial X} \phi(X,t) + \sqrt{2D}\xi(t), \label{EOM} \end{equation}
where $X=X(t)$ denotes the position of the Brownian particle, and $\xi(t)$ denotes a fluctuating term 
that satisfies $\left<\xi(t)\xi(t^{\prime})\right>=\delta(t-t^{\prime})$ with delta function $\delta(\cdot)$.
This novel relationship between the distribution of the upward steps and the active Brownian motion
is analogous to that between the binomial coefficient and Brownian motion.

It can be shown that the distribution $Z(x,t)$ is Gaussian with variance $3Dt/2$ 
under the limit $t\to\infty$ \cite{footnote:Gaussian}.
In that limit, 
although the variance of the OU process (i.e., $\phi(X,t)=X^{2}/2$ in Eq.~(\ref{EOM})) is a constant $D$, 
that of the above time-dependent OU process is proportional to $t$ (Table \ref{table:comparison}).
This is qualitatively the same as the Wiener process (i.e., $\phi(X,t)=0$ in Eq.~(\ref{EOM}));
however, the variance of $Z(x,t)$ is smaller than that of the Wiener process of $2Dt$.
The presence of the potential is included in consideration of the variance.

\begin{table}[!t]%
\caption{Comparison of the variance corresponding to Eq.~(\ref{EOM}).}
\begin{tabular}{lcc}
\hline
\hline
 & ~~potential $\phi(X,t)$~~ & ~~variance ($t\to \infty$)~~ \\
\hline
~~Wiener~~ & $0$ & $2Dt$ \\
~~Ornstein--Uhlenbeck~~ & $X^{2}/2$ & $D$ \\
~~obtained in this Letter~~ & $X^{2}/2t$ & $2Dt/3$  \\
\hline
\hline
\end{tabular}
\label{table:comparison}
\end{table}%

%%%%%%%%%%%%%%%%%%%%%%
{\it A perspective on nonlinear $I$--$V$ property}.---%
Let us leave $Z(x,t)$ with the fictive field $x$ and time $t$ 
and return to $Y(n,N)$ with $n$ intersections of neighboring segmented TDs and array length $N$.
In the long array limit, 
the distribution converges to a Gaussian with variance $N/12$.

Finally, we note the connection of TD to the nonlinearity in the $I$--$V$ behavior.
One can describe the average $I$--$V$ property $I(V) := \overline{I(V, \left\{q\right\})}$, 
where the overline indicates the average for all sets $\{q\}$.
In general, the offset charge distribution affects not only the value of the threshold,
but also the trajectory of the electron between positive and negative electrodes.
Each $I(V, \left\{q_{i}\right\})$ is linear just above its threshold \cite{Bascones2008} as
\begin{equation} I(V,\{q\}) = G(\{q\})(V-V_{\text{th}}(\{q\}))\mathcal{H}(V-V_{\text{th}}(\{q\})), \end{equation}
where $\mathcal{H}(\cdot)$ denotes the Heaviside step function.
The coefficient $G$ depends on the trajectory of an electron and consequently on $\{q\}$.
Here, let us consider 1D arrays,
where $G$ is regarded as a constant for all offset charge distributions;
i.e.,~the offset charge distribution influences only the value of the threshold.
The average $I$--$V$ property of 1D arrays thus reduces to 
$I_{\text{1D}}(V) = \int_{0}^{\infty}I(V,V_{\text{th}})P_{N}(V_{\text{th}}){\rm d}V_{\text{th}}$. 
Further, the conductance reduces to
\begin{equation} \frac{{\rm d} I_{\text{1D}}}{{\rm d}V} = G \int_{0}^{V}P_{N}(V_{\text{th}}){\rm d}V_{\text{th}}, \end{equation}
that is, the conductance is represented by the cumulative distribution of $V_{\text{th}}$.

In the model we employ,
the conductance for long arrays is represented by the error function.
Since it is not unusual that the TD is Gaussian,
a conductance represented by the error function might be universal.
In addition, in higher dimensional arrays, 
we can estimate an approximate $I$--$V$ behavior by a superposition of 1D paths,
although it would be difficult to consider features such as meandering, bifurcation, and confluence.

%%%%%%%%%%%%%%%%%%%%%%
{\it Summary}.---%
We have obtained analytically the TD for a locally-coupled 1D CB array containing $N$ Coulomb islands.
We first found an expression between $V_{\text{th}}$ and $q_{1}$. 
Second, we introduced the segmented TD 
as a sum of products of the probability $\Pi_{N}(k)$ and the conditional probability $U_{N}(n|k)$.
Determining $U_{N}(n|k)$ leads to specific equations for the entire TD
that perfectly describe our simulation results.
In the long-array limit, the distribution converges to Gaussian form with variance $N/12$.
In addition, we discussed a general characteristic of the nonlinear $I$--$V$ behavior,
where the cumulative distribution of the threshold voltage corresponds to the conductance.
The current for each offset charge distribution and confirmation of this viewpoint will be discussed elsewhere.

We also revealed that the distribution of the intersection is equivalent to the distribution $Y(n,N)$,
which indicates the probability for $n-1$ upward steps for $N-1$ aligned objects of different height.
Moreover, the distribution $Z(x,t)$, which is equivalent to $Y(n,N)$, satisfies a novel Fokker--Planck equation 
corresponding to active Brownian motion;
i.e.,~overdamped motion of a Brownian particle in a harmonic potential that spreads with time.
This relationship is analogous to Brownian motion and the binomial coefficients (i.e., the Pascal triangle).
Further, the concept underlying the distribution of upward steps will be applicable to other nonequilibrium and/or disordered systems.
We focused on the derivation of the recurrence formula and the continuous limit in this Letter.
It will be interesting to investigate characteristics of the novel Fokker--Planck equation. 

\begin{acknowledgments}
This work was partially supported by the MEXT, Japan,
a Grant-in-Aid for Scientific Research on Innovative Areas---"Emergence in Chemistry"~(Grant No.~20111003),
and a Grant-in-Aid for Scientific Research (Grant No.~21340110).
\end{acknowledgments}

%%% REFERENCE %%%%%%%%%%%%%%%%%%%%%%%%
%%%%%%%%%%%%%%%%%%%%%%%%%%%%%%%%%%%
%\bibliography{/Users/narupage//Documents/library}

\begin{thebibliography}{99}
\bibitem{footnote:confirmation}%
  We confirmed that $\left<k+1\right>/N=n/N$ is correct for $N\le 6$.
\bibitem{footnote:Gaussian}%
  Let $v_{m}(t)$ denote the $m$-degree moment of $x$ with respect to the the distribution $Z(x,t)$.
  The formal solution of $v_{m}(t)$ is obtained from Eq.~(\ref{PDE_Z}).
  One can derive $v_{m}(t)=0$ for odd $m$ due to symmetry, 
  and $v_{m}(t)=(m-1)!!v_{2}(t)^{\frac{m}{2}}$ ($t\to \infty$) for even $m$.
  It indicates that $Z(x,t)$ converges to a Gaussian distribution.
\end{thebibliography}
%%%%%%%%%%%%%%%%%%%%%%%%%%%%%%%%%%%

%

%%%%%%%%%%%%%%%%%%%%%%%%%%%%%%%%%%%
%%%%%%%%%%%%%%%%%%%%%%%%%%%%%%%%%%%

\end{document}